\documentclass[runningheads]{llncs}
\usepackage[T1]{fontenc}
\usepackage{graphicx,verbatim}
\usepackage{amsmath}
\usepackage{amsfonts}
\usepackage{url}
\usepackage{multirow}
\usepackage{marvosym}

\begin{document}
\title{Case-specific Priors-guided Multimodal Fusion for Prognosis Prediction in Head and Neck Cancer}
\titlerunning{Case-specific Priors-guided Multimodal Fusion for Prognosis Prediction}
\author{
  Yue Lin\inst{1,2}\orcidID{0009-0008-7758-4530}\and
  Shuchang Ye\inst{3}\orcidID{0009-0006-1935-1953}\and
  Shenghai Wu\inst{2}\orcidID{0009-0001-9473-136X}\and
  Dashun Zheng\inst{4}\orcidID{0009-0005-7878-978X}\and
  Mingyuan Meng\inst{1}\orcidID{0000-0002-9562-1613}\Letter
}
\authorrunning{Y. Lin et al.}
\institute{
    Zhongguancun Academy, Beijing, China \and 
    Research Institute, Newland Digital Technology, Fuzhou, China \and 
    School of Computer Science, The University of Sydney, Sydney, Australia 
    \and Faculty of Applied Sciences, Macao Polytechnic University, Macao SAR, China\\
    \email{mengmingyuan@bza.edu.cn}
}
\maketitle

\begin{abstract}
Accurate prognosis prediction for head and neck cancer typically necessitates integrating highly heterogeneous clinical and pathological evidence by sophisticated multimodal fusion. However, existing fusion techniques are usually grounded in case-generic clinical rationales about complementary synergies across modalities, underestimating case-specific variance. In this study, we propose a prognosis prediction framework that explicitly incorporates \textbf{C}ase-specific \textbf{P}riors to guide \textbf{M}ultimodal fusion (named \textbf{CPM}). It leverages broad medical knowledge of LLMs to characterize each case's (case-specific) intra-modal contexts and inter-modal relations, as well as case-generic multimodal rationale, as priors, enabling three-stage priors-guided multimodal fusion: (i) Intra-modal Context-guided Calibration (IntraCC) that leverages case-specific intra-modal priors to guide feature calibration within each modality; (ii) Inter-modal Relation-guided Interaction (InterRI) that leverages case-specific inter-modal priors to guide pair-wise feature interaction in modality pairs; and (iii) Multimodal Rationale-guided Aggregation (MultiRA) that leverages case-generic multimodal priors to guide final feature aggregation over all modalities. On the well-established HANCOCK challenge dataset involving four distinct medical modalities, our CPM achieved a mean AUC of 83.98\% for predicting 5-year overall survival and 2-year recurrence-free survival, improving upon the baseline by 7.58\%, pioneeringly demonstrating the effectiveness of LLM-derived medical priors in guiding multimodal fusion for prognosis prediction. The code is available at \url{https://github.com/yuelin421/CPM}.

\end{abstract}

\section{Introduction}\label{sec:intro}
Head and neck cancer (HNC) remains a major global health burden, with approximately 890,000 new cases and 450,000 deaths worldwide each year \cite{Sung2021CA}. Because patient outcomes vary substantially and clinical decisions rely on risk stratification, accurate prognosis prediction is essential for guiding risk-adapted therapy and follow-up management \cite{Johnson2020NatRevDisPrimers}. To achieve this, prognosis prediction for HNC typically necessitates integrating clinical and pathological evidence via multimodal fusion \cite{Tian2025NPJDM}. However, such evidence is inherently heterogeneous across modalities, posing substantial challenges for multimodal fusion due to ambiguous cross-modal correspondence and mismatched representation granularity \cite{Boehm2022NatRevCancer}.

Early multimodal fusion techniques for prognosis prediction explore cross-modal correspondence in a data-driven manner via end-to-end deep learning. Concretely, many methods adopt generic fusion operators (e.g., attention-based interaction \cite{Chen2021ICCV,Li2023TMI,Meng2024NPJPO} or bilinear-style interaction \cite{Chen2020TMIPathomicFusion}) and rely on survival objectives to induce multimodal synergies during training \cite{Unger2024BMCMedGenomics}. However, these methods failed to optimize multimodal fusion based on valuable clinical rationale, which exists as medical common sense and can provide explicit insight into cross-modal correspondence. To attain clinically grounded multimodal fusion, recent studies have started to develop specialized multimodal fusion techniques based on modality-specific clinical rationale, for example, by designing a PET/CT fusion encoder around the anatomical-metabolic complementarity \cite{Meng2023MICCAI,Lin2025HECKTOR,saeed2026hecktor}, or by introducing pathway-structured representations to mitigate the granularity mismatch between gigapixel histology and global omics profiles \cite{Jaume2024CVPR}. Nevertheless, these designs are inherently coupled to specific modality scenarios and, more importantly, they are anchored to case-generic clinical priors, resulting in limited adaptability to case-specific variance in cross-modal correspondence.

The above limitation suggests a potential direction. If we could obtain case-specific prior knowledge about cross-modal correspondence for each case, multimodal fusion could be guided on a per-case basis rather than being constrained by fixed, case-generic interaction patterns in general clinical rationale. Unfortunately, obtaining case-specific priors at scale remains difficult as it requires expert case-by-case annotation that jointly considers multiple modalities to summarize cross-modal correspondence for each patient. Therefore, such priors are rarely available in existing datasets and have not been explored for multimodal fusion. 
Recently, large language models (LLMs) have made it feasible to derive such priors by automatically analyzing each case’s multimodal evidence and generating clinically plausible summaries based on their broad medical knowledge \cite{Thirunavukarasu2023NatMed,Singhal2025NatMed}. 
Several recent studies have incorporated LLM-derived priors or clinical reports into survival prediction \cite{Kim2024MICCAI,Cui2025MICCAI,Wang2025MICCAI,Zhang2025MICCAI,Liu2025ICLR}. However, they merely exploited LLM to augment the evidence within the modality, where the generated text is treated as either supplementary evidence for a single modality~\cite{Cui2025MICCAI,Wang2025MICCAI,Kim2024MICCAI,Liu2025ICLR}, or as a refined version of the textual modality~\cite{Zhang2025MICCAI}. In contrast, leveraging LLMs to explicitly derive case-specific cross-modal correspondence and further use it to guide multimodal fusion across heterogenous modalities has yet to be explored. 

In this study, we propose CPM, a prognosis prediction framework that explicitly incorporates Case-specific Priors to guide Multimodal fusion. CPM leverages the broad medical knowledge of LLMs to derive case-specific priors that summarize intra-modal contexts and inter-modal relations, together with case-generic multimodal rationale, and then uses them to guide multimodal fusion in a three-stage manner. Specifically, CPM performs priors-guided multimodal fusion via: (i) Intra-modal Context-guided Calibration (IntraCC), which leverages case-specific intra-modal contexts to calibrate each modality stream and aligns modality representations with their corresponding priors through contrastive learning; (ii) Inter-modal Relation-guided Interaction (InterRI), which leverages case-specific inter-modal relations to guide pair-wise feature interaction by conditioning each modality-pair exchange on the relation evidence derived for the current case; and (iii) Multimodal Rationale-guided Aggregation (MultiRA), which leverages the case-generic multimodal rationale to learn rationale-conditioned importance weights over modality pairs and integrates all pairwise interactions into a unified risk representation for final prognosis prediction.

Our CPM was evaluated on the public HANCOCK challenge dataset~\cite{HANCOTHON25GC}, a well-established multimodal HNC dataset with four distinct medical modalities, including (i) whole-slide images (WSIs), (ii) tumor-center tissue microarrays (TMAs), (iii) clinical narratives from medical histories and surgery reports, and (iv) structured clinicopathologic and laboratory measurements \cite{Doerrich2025NatCommun}. Our CPM achieved a mean AUC of 83.98\% for predicting 5-year Overall Survival (OS) and 2-year Recurrence-Free Survival (RFS), improving by 7.58\% over the baseline methods that were not guided by case-specific priors.

\begin{figure}[t]
\centering
\includegraphics[width=\textwidth]{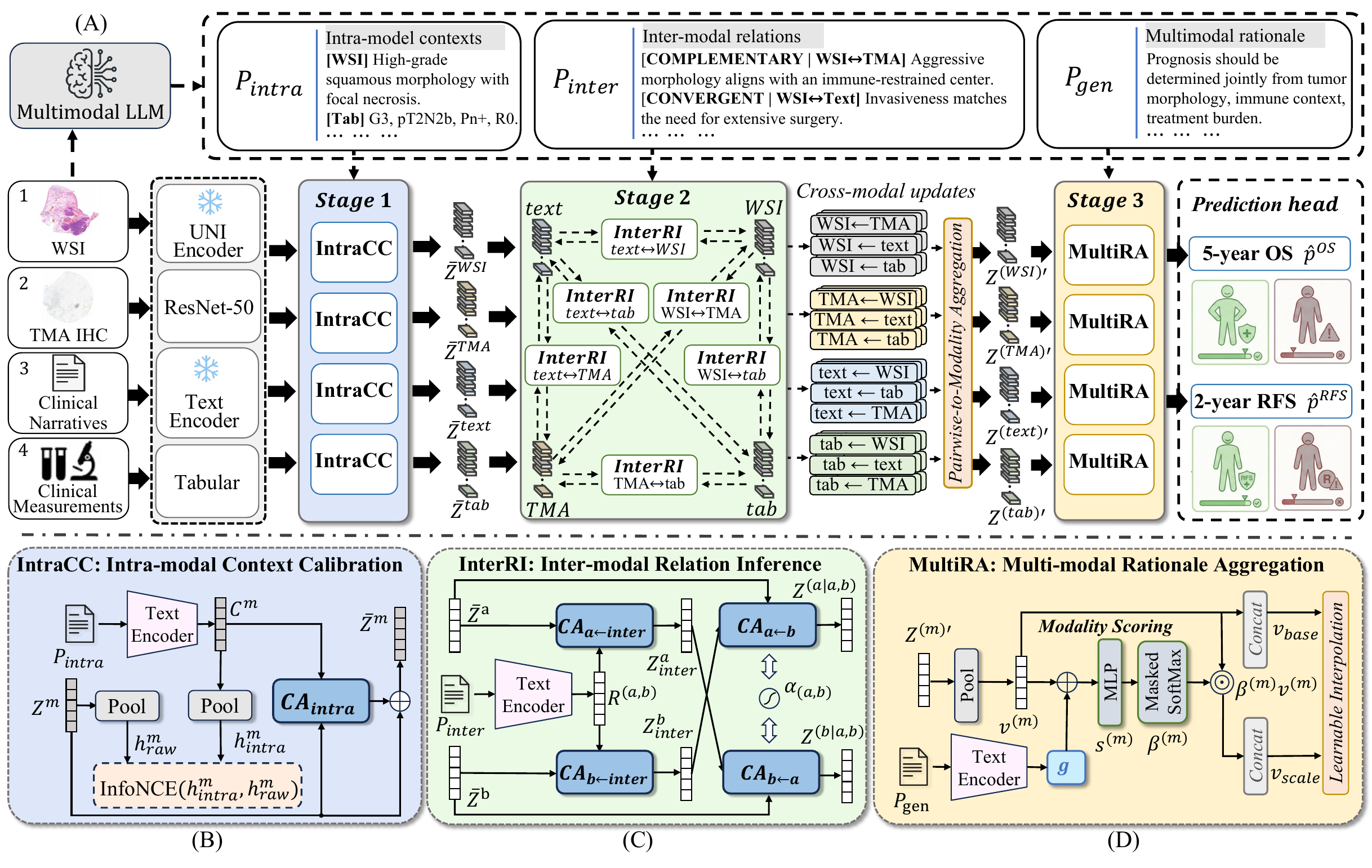}
\caption{\textbf{Overview of CPM.} (A) Overall framework. LLM-derived clinically grounded priors guide multimodal fusion through three stages: (B) IntraCC for intra-modal context calibration, (C) InterRI for relation-guided pairwise interaction, and (D) MultiRA for rationale-guided aggregation.}
\label{fig:framework}
\end{figure}

\section{Methods}
We study multimodal prognosis prediction with \(M\) heterogeneous modalities (\(M=4\) in this study).
For patient \(i\), the model takes the multimodal evidence \(\{x_i^{(m)}\}_{m=1}^{M}\) as input and predicts endpoint probabilities \((\hat{p}_i^{\mathrm{OS}}, \hat{p}_i^{\mathrm{RFS}})\) for OS and RFS, respectively.
We optimize the model with a masked multi-task binary cross-entropy (BCE) objective, where each task loss is computed only when the corresponding label is available.

The overall framework of CPM is shown in Fig.~\ref{fig:framework}(A). 
Given a patient case, we use an LLM to derive clinically grounded priors, as described in Section~\ref{sec:priors}, including case-specific intra-modal contexts, case-specific inter-modal relations, and a case-generic multimodal rationale. These priors then guide fusion in a three-stage pipeline, including Intra-modal Context-guided Calibration (IntraCC, Section~\ref{sec:intraCC}), Inter-modal Relation-guided Interaction (InterRI, Section~\ref{sec:interRI}), and Multimodal Rationale-guided Aggregation (MultiRA, Section~\ref{sec:multiRA}).

\begin{figure}[t]
\centering
\includegraphics[width=\textwidth]{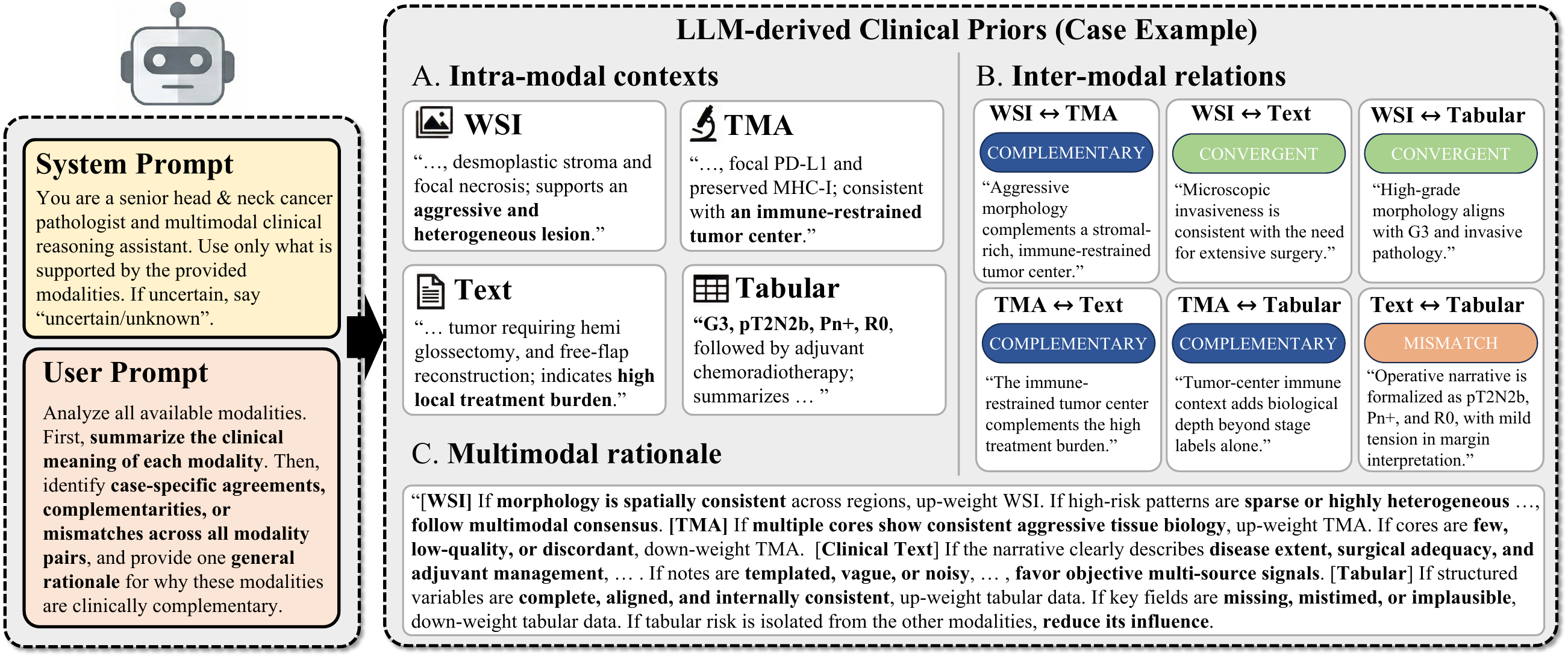}
\caption{Example prompt and output of the LLM-derived clinical priors, including intra-modal contexts, inter-modal relations, and a multimodal rationale.}
\label{fig:prompt}
\end{figure}

\subsection{LLM-derived Clinical Priors}\label{sec:priors}
CPM uses a multimodal LLM to derive clinical priors that guide multimodal fusion with case-specific evidence. 
For each patient \(i\), we ask Qwen3-VL-8B-Instruct\cite{Qwen3VL2025} to conduct pairwise cross-modal assessments over all modality pairs.
Each assessment follows the logic of multidisciplinary case review: it first summarizes intra-modal findings and then characterizes case-level inter-modal agreements or conflicts.
To ensure evidence grounding and consistent formatting, we require anchor-cited statements, allow explicit ``uncertain/unknown'' when evidence is insufficient, and restrict outputs to predefined relation types with confidence scores. 
The final prompt templates, output schema, and LLM generation settings are released with the code to support reproducibility.

From these responses, CPM derives three complementary clinical priors. 
The case-specific intra-modal contexts $\mathcal{P}^{\mathrm{intra}}_i$ provide modality-level LLM interpretations for the case, where each context describes the clinical and pathologic meaning.
The case-specific inter-modal relations $\mathcal{P}^{\mathrm{inter}}_i$ capture pairwise agreements or conflicts to represent cross-modal correspondence within the case.
The case-generic multimodal rationale \(\mathcal{P}^{\mathrm{gen}}\) summarizes clinical rationales about how different modalities complement each other for prognosis prediction. 
Fig.~\ref{fig:prompt} shows representative prompts and outputs.

\subsection{Intra-modal Context-guided Calibration}\label{sec:intraCC}
IntraCC uses the case-specific intra-modal contexts $\mathcal{P}^{\mathrm{intra}}_i$ to calibrate each modality before cross-modal fusion. As shown in Fig.~\ref{fig:framework}(B), for patient $i$ and modality $m$, let $Z_i^{(m)}$ denote modality token embeddings extracted from $x_i^{(m)}$.
The modality-specific context in $\mathcal{P}^{\mathrm{intra}}_i$ is encoded into context token embeddings $C_i^{(m)}$ using a text encoder.
A residual cross-attention block is applied so that modality tokens $Z_i^{(m)}$ can selectively attend to $C_i^{(m)}$. Specifically, cross-attention produces a context-aware update, and residual connection preserves the original modality evidence. As a result, we obtain the calibrated embeddings $\bar{Z}_i^{(m)}$.

To further align raw evidence with the LLM-derived interpretation, an intra-modal contrastive objective is introduced. Pooled embeddings $h_{i,\mathrm{raw}}^{(m)}$ and $h_{i,\mathrm{intra}}^{(m)}$ are obtained from $Z_i^{(m)}$ and $C_i^{(m)}$, respectively. 
Within each batch, we minimize an InfoNCE loss where same-patient pairs are positives and cross-patient pairs are negatives. This loss is computed only for available modalities and added to the masked multi-task BCE loss during training.

\subsection{Inter-modal Relation-guided Interaction}\label{sec:interRI}
InterRI captures case-specific cross-modal correspondence through pairwise interactions guided by the inter-modal relations $\mathcal{P}^{\mathrm{inter}}_i$.
This pairwise design serves as a tractable simplification of full multimodal interaction, since directly characterizing case-specific correspondence across all modalities is highly complex and becomes increasingly difficult as the number of modalities grows, both for modeling and for reliable LLM characterization.
Accordingly, InterRI decomposes the full multimodal interaction into a set of relation-guided pairwise exchanges.

As illustrated in Fig.~\ref{fig:framework}(C), for patient $i$, consider any available modality pair $(a,b)$. The IntraCC-calibrated tokens $\bar{Z}_i^{(a)}$ and $\bar{Z}_i^{(b)}$ are used as inputs and the corresponding relation in $\mathcal{P}^{\mathrm{inter}}_i$ is encoded into relation tokens $R_i^{(a,b)}$.
First, each modality is conditioned on $R_i^{(a,b)}$ via cross-attention so that both modalities are guided by the same pair-specific relation description.
Next, symmetric cross-attention updates both modalities by incorporating relation-conditioned evidence from the other modality.
A learnable pair gate $\alpha_{a,b}$ modulates the interaction strength, and residual connections preserve the IntraCC-calibrated tokens to stabilize fusion.

For each available modality $m$, residual deltas from all relation-conditioned pairs involving $m$ are averaged to form its final update. As a result, the InterRI-updated token embeddings are denoted as $Z_i^{(m)\prime}$.

\subsection{Multimodal Rationale-guided Aggregation}\label{sec:multiRA}
MultiRA aggregates modality representations by using the case-generic clinical rationales in $\mathcal{P}^{\mathrm{gen}}$ to score modalities and scale their representations before fusion.
As shown in Fig.~\ref{fig:framework}(D), for patient $i$ and modality $m$, attentive pooling maps its InterRI-updated tokens $Z_i^{(m)\prime}$ to a modality vector $v_i^{(m)}$. The case-generic clinical rationale in $\mathcal{P}^{\mathrm{gen}}$ are encoded into a shared embedding $g$. 
Let $\mathcal{M}_i$ denote the set of available modalities for patient $i$.
Conditioned on $g$, MultiRA assigns an importance score to each modality and normalizes it across available modalities:
\begin{equation}
s_i^{(m)}=\mathrm{MLP}\!\left([v_i^{(m)}\Vert g]\right),\qquad
\beta_i^{(m)}=\frac{\exp(s_i^{(m)})}{\sum_{k\in\mathcal{M}_i}\exp(s_i^{(k)})}.
\end{equation}

Unlike weighted-sum fusion that collapses all modalities into a single vector, MultiRA scales each modality vector by $\beta_i^{(m)}$ and then concatenates them to form the fused representation. This keeps a dedicated subspace for each modality while allowing the rationale-guided weights to control how much each modality contributes.
To make fusion less sensitive to noisy predicted weights, the scaled concatenation is softly mixed with the unscaled concatenation using a learnable coefficient.
The resulting fused vector is then fed to the prediction head to produce $(\hat{p}_i^{\mathrm{OS}},\hat{p}_i^{\mathrm{RFS}})$.

\section{Experimental Setup}
\subsection{Dataset}
CPM was evaluated on the public HANCOCK challenge dataset~\cite{Doerrich2025NatCommun,HANCOTHON25GC}, which comprises 764 HNC patients and provides four distinct medical modalities. 
WSIs capture fine-grained tumor morphology, and HANCOCK provides pre-extracted UNI patch embeddings~\cite{Chen2024NM}. 
In parallel, TMA cores with immunohistochemistry staining capture immune and tumor phenotypes.
These pathology modalities are complemented by clinical narratives that summarize patient history and treatments, together with structured clinicopathologic and laboratory measurements that reflect disease characteristics. All experiments followed the official HANCOCK challenge split, comprising 611 training patients and 152 test patients.

\subsection{Implementation Details}
\textbf{Training scheme.}
CPM was implemented in PyTorch and trained on 2 NVIDIA GeForce RTX 5090 GPUs for 50 epochs with a batch size of 12 using the AdamW optimizer. 
We set the learning rate to \(1\times10^{-5}\) and used a weight decay of \(1\times10^{-2}\). 
The objective combined a masked multi-task BCE loss with an InfoNCE loss used in IntraCC, weighted by 0.1 and using a temperature of 0.07. Model selection was performed on the training split via 5-fold cross-validation, where checkpoints were chosen by the best mean validation AUC across folds.

\noindent\textbf{Feature extraction.}
For WSIs, we directly used the pre-extracted UNI patch embeddings provided by HANCOCK~\cite{Chen2024NM,Doerrich2025NatCommun}. 
For TMAs, we resized each core image to \(512\times512\) and encoded it with an ImageNet-pretrained ResNet-50 backbone.
For clinical narratives, we concatenated them and encoded with ClinicalBERT~\cite{alsentzer2019clinicalbert} using a maximum length of 512, and froze the text encoder during training. 
For tabular data, we standardized numerical variables by z-score normalization and encoded categorical variables by one-hot encoding to construct tabular features. 
Finally, we tokenized the LLM-derived clinical priors with a maximum length of 512 and encoded them with the same frozen ClinicalBERT to obtain prior features that guide IntraCC, InterRI, and MultiRA.

\subsection{Experiment Designs}
Comparisons were designed to cover the two main fusion paradigms discussed in Section~\ref{sec:intro}, including data-driven and clinically grounded fusion.
As reference baselines, we reported unimodal results for each modality to quantify standalone performances, and included a simple feature concatenation model as well as a Random Forest classifier to contextualize multimodal gains.
We further included recent LLM-assisted prognosis methods for comparison.
For all compared methods, we reused the same extracted modality features described above to ensure a controlled comparison.
Top-performing methods from HANCOCK challenge were not included since they were not publicly released and the hidden test set is not accessible for direct evaluation.\footnote{\url{https://hancothon25.grand-challenge.org/}}

We also performed stage-wise ablations by removing one stage of CPM at a time to evaluate the contribution of each stage. Statistical significance of pairwise AUC differences between methods was assessed using DeLong's test.

\section{Results and Discussion}
Table~\ref{tab:main} reports the comparison results on HANCOCK.
Using the tabular modality alone outperforms several multimodal methods, suggesting that effective fusion across the four distinct modalities is highly challenging.
This result indicates that many existing fusion methods fail to fully exploit cross-modal synergy in HANCOCK.
Data-driven fusion methods rely mainly on end-to-end supervision to discover multimodal correspondence, but they do not leverage valuable clinical priors.
Clinically grounded fusion methods incorporate prior knowledge, yet their designs are tailored to modality-specific scenarios, which limits their adaptability in this setting.
LLM-assisted methods mainly use LLM outputs as supplementary evidence for individual modalities, but they do not explicitly exploit case-specific cross-modal correspondence for multimodal fusion.

By contrast, CPM explicitly derives case-specific priors to characterize intra-modal contexts and inter-modal relations, together with a case-generic multimodal rationale, and uses them to guide multimodal fusion through IntraCC, InterRI, and MultiRA.
This design enables CPM to more effectively uncover complementary synergies across the four HANCOCK modalities and thereby achieve the best overall performance, surpassing both unimodal settings and existing multimodal methods.

Table~\ref{tab:ablation} reports the ablation results of CPM.
Removing IntraCC, InterRI, or MultiRA consistently reduces the overall AUC, confirming that each stage contributes to the effectiveness of CPM.
Even after removing a single stage, the results remain competitive with representative methods in Table~\ref{tab:main}, supporting the value of case-specific LLM-derived priors for guiding multimodal fusion.

\begin{table}[t]
\caption{\textbf{Comparison on HANCOCK.} Best results are highlighted in bold. * indicates \(p<0.05\) for AUC compared with CPM using DeLong's test.}
\label{tab:main}
\centering
\footnotesize
\setlength{\tabcolsep}{3.6pt}
\renewcommand{\arraystretch}{1}
\begin{tabular}{lcc|cc|cc}
\hline
\multirow{2}{*}{Method} &
\multicolumn{2}{c|}{5-year OS} &
\multicolumn{2}{c|}{2-year RFS} &
\multicolumn{2}{c}{Average} \\
& AUC & Acc & AUC & Acc & AUC & Acc \\
\hline
\multicolumn{7}{l}{\textbf{Reference baselines}} \\
WSI-only      & 0.7648$^{*}$ & 0.6479 & 0.7049$^{*}$ & 0.7593 & 0.7349 & 0.7036 \\
TMA-only      & 0.6573$^{*}$ & 0.4444 & 0.6309$^{*}$ & 0.6818 & 0.6441 & 0.5631 \\
Text-only     & 0.6301$^{*}$ & 0.5417 & 0.6935$^{*}$ & 0.7364 & 0.6618 & 0.6391 \\
Tabular-only  & 0.7615$^{*}$ & 0.6528 & 0.7782$^{*}$ & 0.7545 & 0.7699 & 0.7037 \\
Random Forest & 0.7078$^{*}$ & 0.6389 & 0.7248$^{*}$ & 0.7364 & 0.7163 & 0.6877 \\
Concat + MLP  & 0.7695$^{*}$ & 0.7465 & 0.7580$^{*}$ & \textbf{0.7870} & 0.7640 & 0.7668 \\
\hline
\multicolumn{7}{l}{\textbf{Data-driven fusion methods}} \\
MCAT~\cite{Chen2021ICCV}
              & 0.6754$^{*}$ & 0.6338 & 0.7795$^{*}$ & 0.7315 & 0.7275 & 0.6827 \\
MOTCat~\cite{Xu2023ICCV_MOTCat}
              & 0.8070$^{*}$ & 0.6901 & 0.7080$^{*}$ & 0.6759 & 0.7575 & 0.6830 \\
\hline
\multicolumn{7}{l}{\textbf{Clinically grounded fusion methods}} \\
SurvPath~\cite{Jaume2024CVPR}
              & 0.8101$^{*}$ & 0.7165 & 0.6942$^{*}$ & 0.7407 & 0.7522 & 0.7286 \\
\hline
\multicolumn{7}{l}{\textbf{LLM-assisted methods}} \\
TMSE~\cite{Zhang2025MICCAI}
              & 0.7959$^{*}$ & 0.6901 & 0.6737$^{*}$ & 0.7407 & 0.7348 & 0.7154 \\
LLM-guided MIL~\cite{Kim2024MICCAI}
              & 0.7368$^{*}$ & 0.6901 & 0.7852$^{*}$ & 0.7200 & 0.7610 & 0.7051 \\
CPM (Ours)    & \textbf{0.8461} & \textbf{0.7746} & \textbf{0.8335} & \textbf{0.7870} & \textbf{0.8398} & \textbf{0.7808} \\
\hline
\end{tabular}
\end{table}

\begin{table}[tb]
\caption{\textbf{Ablations on HANCOCK.} Best results are highlighted in bold. * indicates \(p<0.05\) for AUC compared with CPM using DeLong's test.}
\label{tab:ablation}
\centering
\footnotesize
\setlength{\tabcolsep}{3.6pt}
\renewcommand{\arraystretch}{1}
\begin{tabular}{lcc|cc|cc}
\hline
\multirow{2}{*}{Variant} &
\multicolumn{2}{c|}{5-year OS} &
\multicolumn{2}{c|}{2-year RFS} &
\multicolumn{2}{c}{Average} \\
& AUC & Acc & AUC & Acc & AUC & Acc \\
\hline
CPM (Ours)    & \textbf{0.8461} & \textbf{0.7746} & \textbf{0.8335} & \textbf{0.7870} & \textbf{0.8398} & \textbf{0.7808} \\
w/o IntraCC   & 0.8381$^{*}$ & 0.7606 & 0.8013$^{*}$ & 0.7315 & 0.8197 & 0.7461 \\
w/o InterRI   & 0.8094$^{*}$ & 0.7324 & 0.8299$^{*}$ & \textbf{0.7870} & 0.8196 & 0.7597 \\
w/o MultiRA   & 0.8134$^{*}$ & \textbf{0.7746} & 0.8116$^{*}$ & \textbf{0.7870} & 0.8125 & \textbf{0.7808} \\
\hline
\end{tabular}
\end{table}

\section{Conclusion}
We presented CPM, a case-specific priors-guided multimodal prognosis prediction framework that uses LLM-derived case-specific priors to guide multimodal fusion for HNC. By modeling intra-modal contexts, inter-modal relations, and multimodal rationales, CPM performs priors-guided multimodal fusion through IntraCC, InterRI, and MultiRA. Experimental results on HANCOCK demonstrated that CPM consistently outperformed representative baselines and competing methods, highlighting the effectiveness of explicitly modeling case-specific cross-modal correspondence for multimodal prognosis prediction.

\medskip
\noindent\textbf{Acknowledgments.}
This work was supported by the Zhongguancun Academy (Grant No. XTS0071).

\medskip
\noindent\textbf{Disclosure of Interests.}
The authors have no competing interests to declare.

\bibliographystyle{splncs04}
\bibliography{mybibliography}
\end{document}